\documentclass[conference]{IEEEtran}
 
\usepackage{setspace,amssymb,graphicx,enumerate,amsfonts,amsmath,color,algorithm,algpseudocode}
\usepackage{epsfig,subfigure,booktabs,amsmath,calc,arydshln,tikz,pdflscape,rotating}
\usepackage{cite}
\usepackage[english]{babel}
\usepackage{epstopdf,multirow}
\usetikzlibrary{shapes}
\usepackage{calc}
\usepackage{ifthen}
\usepackage{verbatim}
\usepackage{multicol}

\IEEEoverridecommandlockouts

\begin{document}
\bstctlcite{IEEEexample:BSTcontrol}

\title{Energy Efficiency Optimization of Channel Access Probabilities in IEEE 802.15.6 UWB WBANs}
\author{\IEEEauthorblockN{Yang Liu, Kemal Davaslioglu, and Richard D. Gitlin}
\IEEEauthorblockA{Department of Electrical Engineering, University of South Florida\\
Email: yangl@mail.usf.edu, \{kemald, richgitlin\}@usf.edu}}
\maketitle
\begin{abstract}
Energy efficiency is essential for Wireless Body Area Network (WBAN) applications because of the battery-operated nodes. Other requirements such as throughput, delay, quality of service, and security levels also need to be considered in optimizing the network design. In this paper, we study the case in which the nodes access the medium probabilistically and we formulate an energy efficiency optimization problem under the rate and access probability constraints for IEEE 802.15.6 Impulse Radio Ultra-wideband (IR-UWB) WBANs. The proposed algorithm, dubbed Energy Efficiency Optimization of Channel Access Probabilities (EECAP), determines the optimal access probability and payload frame size for each node. The simulation results show that our algorithm rapidly converges to the optimal solution. We also provide detailed insights on the relationship between the optimal access probabilities and other network parameters such as the link distance, the number of nodes, and the minimum rate constraints.
\end{abstract}

\section{Introduction}

Wireless Body Area Networks (WBANs) connect wearable computing devices and sensors that are placed on, around or inside the body through wireless networks. They enable many promising applications in the area of remote health monitoring, health care, fitness, smart clothing, etc. To address the unique demands of WBANs, the IEEE 802.15.6 standard was proposed and finalized in 2012 \cite{IEEE802156}. Operating in three different modes including narrowband, ultra-wideband (UWB) and human body communications, IEEE 802.15.6 standardizes the physical layer (PHY) and a common medium access control (MAC) layer protocol. We focus on the UWB mode in this paper, since it offers more robustness against channel variations, transmits at ultra-low power, and achieves high data rates for human body applications \cite{IEEE802156,Witrisal09}.

Energy efficiency is of vital importance for WBANs because most of the wearable devices or sensors are battery-operated. For this reason, energy efficiency for WBANs has been well studied in the literature, see, e.g., \cite{Franco10,Mohammadi14,Mohammadi14b,Karvonen14,CLOEE16}. In \cite{Franco10}, the authors present an analytical model for estimating the device lifetime and evaluate the energy lifetime performance for contention free access. The length of the MAC frame body is optimized in \cite{Mohammadi14} to maximize the energy efficiency in IEEE 802.15.6 UWB WBANs. A novel link adaptation strategy that aims to maximize energy efficiency for IEEE 802.15.6 IR-UWB systems is proposed in \cite{Mohammadi14b}.

Both random access and scheduled access mechanisms are defined in the standard \cite{IEEE802156}. We consider the distribution of the nodal channel random access probabilities, which can have a large impact on the network’s throughput, delay, as well as energy efficiency. In \cite{Karvonen14}, a cross-layer energy efficiency optimization model that includes a probability of success for the MAC layer is studied. The model takes the channel access probabilities into account. However, this study does not provide an optimal distribution of access probabilities for multiple sensors and does not consider two or more parameters in their optimization. To address these shortcomings, in this paper, we formulate an energy efficiency maximization problem for the IEEE 802.15.6 UWB WBANs. Building on our prior work in \cite{CLOEE16} that determines the optimal frame size and modulation scheme for WBANs, we now extend the framework to include random channel access probabilities. We derive the MAC layer successful transmission probabilities. The proposed algorithm, Energy Efficiency Optimization of Channel Access Probabilities (EECAP), determines the optimal channel access probability and frame length that maximize the energy efficiency of multiple sensors under the rate and access constraints. 


The rest of this paper is organized as follows. Section \ref{Section:SystemModel} introduces the IEEE 802.15.6 WBAN system model, network topology, and the PHY and MAC layer parameters. The error probabilities in each layer are derived. In Section \ref{Section:Formulation}, we formulate problem and describe the EECAP algorithm. In Section \ref{Section:Simulations}, we present the simulation results. Finally, concluding remarks are made in Section \ref{Section:Conclusion}.

\section{System Model}\label{Section:SystemModel}
In this section, we introduce the IEEE 802.15.6 UWB PHY and MAC layer parameters \cite{IEEE802156} that are employed in this paper. Due to the space considerations, we briefly discuss these parameters and refer the reader to \cite{IEEE802156} and Section II in \cite{CLOEE16} for details. Also, we derive the probability of error in the PHY layer. For the MAC layer, the successful transmission, collision, and idle channel probabilities are obtained. Finally, we describe the time duration and energy consumption models considered in this paper.
\subsection{IEEE 802.15.6 UWB PHY Superframe Structure}\label{Subsection:TwoA}
\begin{figure}[t!]
\centering
\includegraphics[width=0.8\columnwidth]{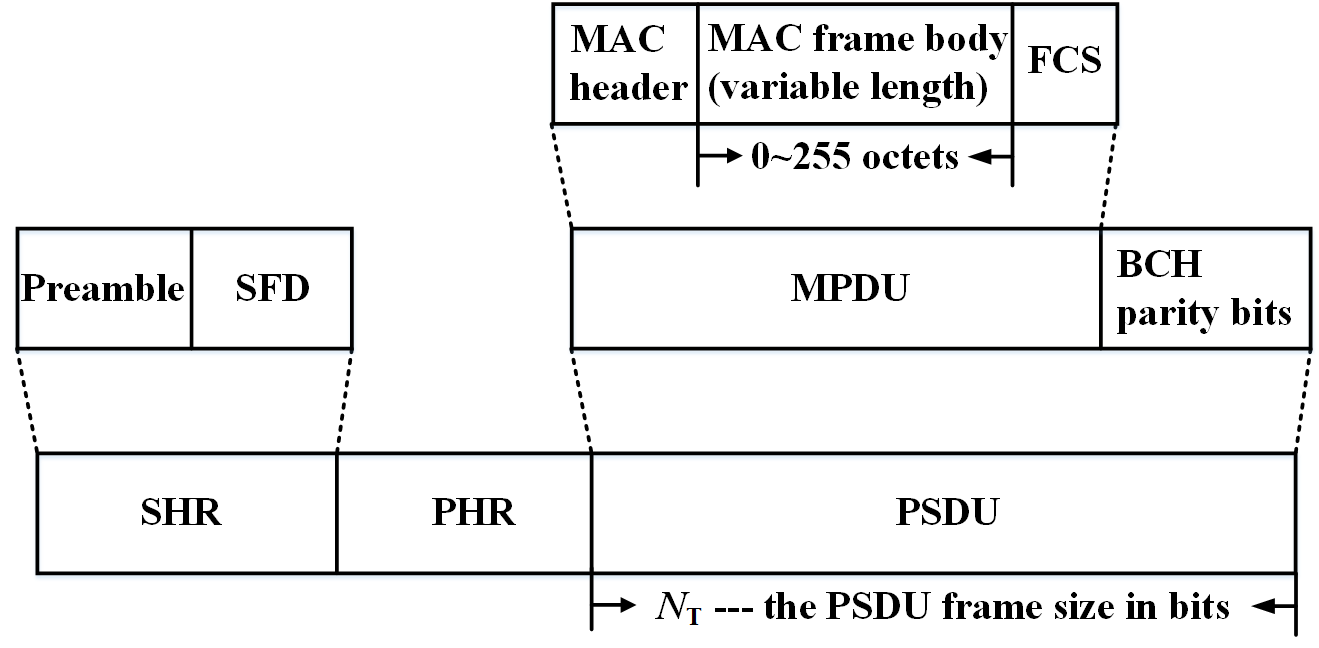}
\caption{IEEE 802.15.6 UWB PPDU superframe structure.}
\label{Fig1}  
\end{figure}

The format of IEEE 802.15.6 UWB PHY superframe is depicted in Fig.~\ref{Fig1}. The entire superframe is referred as the physical layer protocol data unit (PPDU) and it consists of the synchronization header (SHR), the physical layer header (PHR), and the physical layer service data unit (PPDU). We denote the length of the PSDU frame after channel encoding as $N_{\text{T}}$, which is one of our optimization parameters. 

The time duration of the superframe is given by
\begin{align} 
T_{\text{PPDU}} = T_{\text{SHR}} + T_{\text{PHR}} + T_{\text{PSDU}},
\end{align}
where $T_{\text{SHR}}$, $T_{\text{PHR}}$, and $T_{\text{PSDU}}$ stand for the time durations for the SHR, PHR, and PSDU, respectively. The values of $T_{\text{SHR}}$ and $T_{\text{PHR}}$ are given by the standard \cite{IEEE802156}, $T_{\text{SHR}}=40.32$ $\mu$sec and $T_{\text{PHR}}=80.052$ $\mu$sec. The value of $T_{\text{PSDU}}$ changes with the variable MAC frame body, $T_{\text{PSDU}}=N_{\text{T}}T_{\text{sym}}$, where $T_{\text{sym}}$ is the symbol duration. The Bose-Chaudhuri-Hocquenghem (BCH) channel code $(n=63,k=51;t=2)$ is used \cite{IEEE802156}, where $k$, $n$, and $t$ denote the message, codeword, and error correction capability in bits, respectively. After channel encoding, the number of codewords in the PSDU frame is \cite{CLOEE16}
\begin{align} 
N_{\text{CW}}=\left\lceil\frac{8N_{\text{MACframebody}}+72}{k}\right\rceil=\frac{N_{\text{T}}}{n},
\end{align}
where $N_{\text{MACframebody}}$ is the number of octets in the variable MAC frame body.

\subsection{Error Corrections}\label{Subsection:TwoB}
For a non-coherent energy detection (ED) receiver, the bit error probability for IR-UWB transceivers \cite{Witrisal09,Karvonen14} is given by
\begin{align}
P_b = Q\left(\sqrt{\frac{1}{2} \cdot \frac{(h\varepsilon_b/N_0)^2}{h \varepsilon_b / N_0 + N_{\text{cpb}} T_{\text{int}} W_{\text{rx}}}}\right),
\end{align}
where $h$ is the channel coefficient, $N_0$ is the noise power level, and $ W_{rx} $ is the equivalent noise bandwidth of the receiver front end. The number of pulses per burst is denoted as $ N_{\text{cpb}} $, which is one of the optimized parameters in our previous work \cite{CLOEE16}. For the single pulse option, $ N_{\text{cpb}}=1 $ and for the burst pulse option, $ N_{\text{cpb}}\in\{2,4,8,16,32\} $. This parameter is used to balance the trade-off between the processing gain and the symbol rate. The integration interval per pulse is denoted by $ T_{int} $ and we select $ T_{int}=N_{\text{cpb}}T_p $, where $ T_p $ is the time duration of a single pulse. The energy of a burst is $ \varepsilon_b $. The energy of a pulse is $ \varepsilon_p=\varepsilon_b/N_{\text{cpb}} $ and the signal to noise ratio per bit can be expressed as $ \varepsilon_p/N_0 $.

In the following, we derive the error probability of the SHR, PHR, and PSDU frames. 
In the SHR frame, a 63-bit Kasami sequence is used in the  start-of-frame delimiter (SFD) and four Kasami sequences are used in the preamble. The probability of correct decoding of one Kasami sequence is \cite{CLOEE16}
\begin{align}
P_{\text{Kasami}} = \sum_{i = 0}^{\rho} \binom{63}{i} (P_b)^i (1-P_b)^{63 - i},
\end{align}
where $ \rho $ stands for an implementation-dependent sensitivity margin and is taken as $ \rho=6 $ as in \cite{CLOEE16}. Therefore, the probability of successfully receiving the SHR frame is 
\begin{align}
P_{\text{SHR}}=P_{\text{Preamble}}P_{\text{SFD}},
\end{align}
where $ P_{\text{Preamble}}=1-(1-P_{\text{Kasami}})^4 $ and $ P_{\text{SFD}}=P_{\text{Kasami}}$.
Next, for the PHR frame, the probability of correct reception is 
\begin{align}
P_{\text{PHR}}  = \sum_{i = 0}^{t} \binom{N_{\text{PHR}}}{i} (P_b)^i (1-P_b)^{N_{\text{PHR}} - i},
\end{align}
where $ N_{\text{PHR}}=40 $ is the number of bits in the PHR frame and $ t_{\text{ECC}}=2 $ is the error correcting capability of the (40,28;2) BCH code that is used in the PHR frame. Finally, there are $ N_{\text{CW}} $ codewords in the PSDU frame. The probability of receiving the PSDU frame successfully can be expressed as
\begin{align}
P_{\text{PSDU}}=(P_{\text{CW}})^{N_{\text{CW}}}=(P_{\text{CW}})^{\frac{N_{\text{T}}}{n}},
\end{align} 
where $ P_{\text{CW}} $ represents the probability of receiving one codeword correctly, which is given by
$P_{\text{CW}} = \sum_{i = 0}^{t} \binom{n}{i} (P_b)^i  (1-P_b)^{n - i}$.
Hence, the probability of successful reception of the PPDU frame can be expressed as 
\begin{align}
P_{\text{PPDU}}  = P_{\text{SHR}}  P_{\text{PHR}}  P_{\text{PSDU}} = P_{\text{SHR}}  P_{\text{PHR}} ( P_{\text{CW}} )^{\frac{N_{\text{T}}}{n}}.
\end{align}
On one hand, we have $P_{\text{PPDU}}$ decrease as $ N_{\text{T}} $ increases for $ P_{\text{CW}}<1 $. On the other hand, as $ N_{\text{T}} $ increases, the proportion of the overhead in the superframe decreases, which results in a higher system efficiency.

\subsection{Network Topology and Channel Access Probabilities}\label{Subsection:TwoC}
\begin{figure}[t!]
	\centering
	\includegraphics[width=0.45\columnwidth]{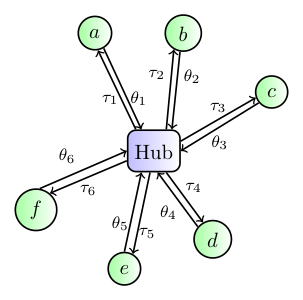}
	\caption{One-hop WBAN star network topology consisting of multiple nodes and a hub. Nodes send their application requirements and the hub sends back their channel access probabilities.}
	\label{Fig2}  
\end{figure}

As shown in Fig.~\ref{Fig2}, we consider a one-hop star topology WBAN, which has one hub and $ N_{S} $ nodes. For node $ k $, $ k=1,2,...,N_{S} $, the channel access probability, PSDU frame body size, distance from the hub and its minimum rate constraint are denoted as $ \tau_k $, $ N_k^{\text{T}} $, $ d_k $, and $ R_k^{\min} $, respectively. The scenario we investigate is that every node contends for the medium and the hub works as a controller to determine the optimal access probability of the nodes. Fig.~\ref{Fig2} depicts an example scenario where multiple nodes report their requirements, denoted by $ \theta_k $, and the hub assigns their channel access probabilities. The node requirements can be multi-bit signals, and are application and node specific addressing a variety of constraints such as the rate, delay, power, reliability, QoS, and security levels. In this paper, we only focus on the minimum rate constraint for simplicity, i.e., $ \theta_k=R_k^{\min} $, although the proposed framework can address multiple constraints at the same time.

Similar to the channel states described in \cite{Bianchi00,Giarre09} for IEEE 802.11 networks, we define three channel states for the one-hop star WBAN: 1). Successful transmission: one of the nodes gets the channel and successfully transmits its packets; 2). Collision: more than one user transmit packets and they collide; 3). Idle channel: none of the nodes transmits.

For node $ k $, the probability of successful transmission is
\begin{align}
P_k^{\text{S}}=\tau_k\prod_{j \neq k}\big(1-\tau_j)=\tau_k(1-p_k),
\end{align}
where $ p_k=1-\prod_{j \neq k}(1-\tau_j) $ is the collision probability experienced by node $ k $ because of other nodes \cite{Giarre09}. Then, the probability of successful transmission for all the nodes is the sum of each node's probability of success,
\begin{align}\label{eq11}
P^{\text{S}}=\sum_{k}P_k^{\text{S}}=\sum_{k}\tau_k\big(1-p_k).
\end{align}
The probability of an idle channel can be expressed as $P^{\text{I}}=\prod_{k}\big(1-\tau_k)$. Then, the probability of collision is given by $P^{\text{C}}=1-P^{\text{S}}-P^{\text{I}}$. Thus, the above probabilities can be expressed by linear functions of $ \tau_k $, where $ k=1,2,...,N_{\text{S}} $, as
\begin{align}
P^{\text{S}}=x_k^{\text{S}}\tau_k + y_k^{\text{S}}, P^{\text{C}}=x_k^{\text{C}}\tau_k + y_k^{\text{C}}, P^{\text{I}}=x_k^{\text{I}}\tau_k + y_k^{\text{I}},
\end{align}
where $ x_k^{\text{C}}=\sum_{j \neq k}\tau_j\frac{1-p_j}{1-\tau_k} $, $ x_k^{\text{S}}=1-p_k-x_k^{\text{C}} $, $ x_k^{\text{I}}=p_k-1 $ and $ y_k^{\text{S}}=x_k^{\text{C}} $, $ y_k^{\text{C}}=p_k-x_k^{\text{C}} $, $ y_k^{\text{I}}=1-p_k $. These terms will help us obtain the closed-form expressions in Section \ref{Section:Formulation}.

\subsection{Time Duration Model}\label{Subsection:TwoD}
The time duration will also be different for each of the three channel states. If node $ k $ successfully transmits its packet and receives the acknowledgment packet from the hub, the time duration is given by
\begin{align}
T_k^{\text{S}}=T_k^{\text{PPDU}}+T_{\text{ACK}}+2T_{\text{pSIFS}}+2\sigma_k,
\end{align}
where $ T_k^{\text{PPDU}}=T_{\text{SHR}}+T_{\text{PHR}}+N_k^{\text{T}}T_{sym} $ is node $ k $'s PPDU time duration, which has been defined in Section \ref{Subsection:TwoA}. The PSDU frame size for node $ k $ is denoted as $ N_k^{\text{T}} $. The time duration of the acknowledgment packet from the hub is $T_{\text{ACK}}=T_{\text{SHR}}+T_{\text{PHR}}+N_{\text{T}}^{\min}T_{sym} $. We assume it uses the minimum frame length, $ N_{\text{T}}^{\min}=126 $ bits \cite{IEEE802156}. The time period of the short interframe spacing is $ T_{\text{pSIFS}}=75 $ $ \mu $sec \cite{IEEE802156}. The propagation time is denoted by $ \sigma_k $. The time spent in the scenario of collision is given by
\begin{align}
T_k^{\text{C}}=T_k^{\text{PPDU}}+T_{\text{pSIFS}}+\sigma_k.
\end{align}
The time period in an idle channel is given by the CSMA slot time period. From \cite{IEEE802156}, we have $ T_k^{\text{I}}=292 $ $ \mu $sec. 

\subsection{Energy Consumption Model}\label{Subsection:TwoE} 
Among the energy models for IR-UWB radios \cite{ Mohammadi14,Karvonen14,Seyedi10}, we employ the one in \cite{Seyedi10} as it provides a general model which considers different detectors, modulation types and demodulators. We define the energy consumptions for each of the three channel states described in Section~\ref{Subsection:TwoC}. For a successful transmission, the energy consumption is 
\begin{align}
\varepsilon_k^{\text{S}}=\varepsilon_{\text{B}}N_k^{\text{T}}+\varepsilon_{\text{OH}}+\varepsilon_{\text{ST}},
\end{align}
where $ \varepsilon_{\text{B}} $ stands for the energy required to transmit and receive a payload bit, $ \varepsilon_{\text{OH}} $ is the energy consumption for the transmission and reception of the overhead, and $ \varepsilon_{\text{ST}} $ denotes the startup energy. When collision happens, the energy is given by
\begin{align}
\varepsilon_k^{\text{C}}=\varepsilon_{\text{B}}^{\text{Tx}}N_k^{\text{T}}+\varepsilon_{\text{OH}}^{\text{Tx}}+\varepsilon_{\text{ST}}^{\text{Tx}},
\end{align}
where $ \varepsilon_{\text{B}}^{\text{Tx}} $, $ \varepsilon_{\text{OH}}^{\text{Tx}}$, and $ \varepsilon_{\text{ST}}^{\text{Tx}} $ are defined similarly as the energy required for payload bit, overhead, and startup in transmission only. When the channel is idle, we assume that no energy is consumed, i.e., $ \varepsilon_k^{\text{I}}=0 $ \cite{Seyedi10}.
The details about the energy consumption model for different detector, modulation, and demodulation types can be found in \cite{CLOEE16,Seyedi10}.

\section{Energy Efficiency Maximization Problem}\label{Section:Formulation}
The energy efficiency and throughput are defined to aid the problem formulation:
\newtheorem{definition}{Definition}
	\begin{definition}
	The energy efficiency for node $ k $ is defined as the successfully transmitted payload information divided by the average energy consumption, which can be expressed as
		\begin{align}
		\eta_k = \frac{N_k^{\text{T}}P_k^{\text{S}}P_k^{\text{SHR}}P_k^{\text{PHR}}(P_k^{\text{CW}})^{\frac{N_k^{\text{T}}}{n}}}{P^{\text{S}}\varepsilon_k^{\text{S}}+P^{\text{C}}\varepsilon_k^{\text{C}}+P^{\text{I}}\varepsilon_k^{\text{I}}}.
		\end{align}	
	\end{definition}
\begin{definition}
	Node $ k $'s throughput is defined as the number of successfully transmitted payload bits divided by the average time duration, which is given by
		\begin{align}
		R_k = \frac{N_k^{\text{T}}P_k^{\text{S}}P_k^{\text{SHR}}P_k^{\text{PHR}}(P_k^{\text{CW}})^{\frac{N_k^{\text{T}}}{n}}}{P^{\text{S}}T_k^{\text{S}}+P^{\text{C}}T_k^{\text{C}}+P^{\text{I}}T_k^{\text{I}}}.
	\end{align}	
\end{definition}
\newtheorem{lemma}{Lemma}
\begin{lemma}
	The derivative of the node $ k $'s throughput with respect to $ \tau_k $ is a monotonically increasing function, i.e., $ \partial{R_k}/\partial \tau_k \geq 0 $, which is proved in Appendix. Consequently, the minimum value of $ \tau_k $ that satisfies the rate constraint can be obtained by letting $ R_k=R_k^{\min} $ and rearranging terms as
	\begin{align}\label{taukmin_THR}
	\tau_{k,\min}^{\text{THR}}=\frac{R_k^{\min} \cdot YT}{N_k^{\text{T}}(1-p_k)P_k^{\text{SHR}}P_k^{\text{PHR}}(P_k^{\text{CW}})^{\frac{N_k^{\text{T}}}{n}}-R_k^{\min} \cdot XT},
	\end{align}
	where $ XT=x_k^{\text{S}}T_k^{\text{S}}+x_k^{\text{C}}T_k^{\text{C}}+x_k^{\text{I}}T_k^{\text{I}} $ and $ YT=y_k^{\text{S}}T_k^{\text{S}}+y_k^{\text{C}}T_k^{\text{C}}+y_k^{\text{I}}T_k^{\text{I}} $. If $ \tau_{k,\min}^{\text{THR}} \notin (0,1) $, there is no feasible solution for $ \tau_k $ that satisfies the rate constraint.
\end{lemma}
\begin{lemma}
	If $ P_k^{\text{CW}} \in (0,1) $ holds, then the optimal PSDU frame length for the node $ k $'s throughput is obtained by letting $ \partial R_k/\partial N_k^{\text{T}}=0 $, and its closed-form expression is 
	\begin{align}\label{NkT_THR}
	N_{k,\text{T}}^{\text{THR}} = \left[ -\frac{\left[n+\log(P_k^{\text{CW}})\right]\cdot TO}{\log(P_k^{\text{CW}}) \cdot TN} \right]_{N_{\text{T}}^{\min}}^{N_{\text{T}}^{\max}},
	\end{align}
	where $ TO=P^{\text{S}}(T_{\text{SHR}}+T_{\text{PHR}}+T_{\text{ACK}}+2T_{\text{pSIFS}}+2\sigma_k)+P^{\text{C}}(T_{\text{SHR}}+T_{\text{PHR}}+T_{\text{pSIFS}}+\sigma_k)+P^{\text{I}}T_k^{\text{I}} $ and  $ TN=(P^{\text{S}}+P^{\text{C}})T_{sym} $. The notation $ \left[x\right]_a^b $ denotes that $x$ is lower bounded by $a$ and upper bounded by $b$. If $ P_k^{\text{CW}}=0 $, then the throughput is always zero. If $ P_k^{\text{CW}}=1 $, then $N_{k,\text{T}}^{\text{THR}}=N_{\text{T}}^{\max} $. 
\end{lemma}

The problem \textbf{(EE)} maximizes the network energy efficiency subject to the minimum rate constraint and the access probability constraint, which can be formulated as
\begin{subequations}\label{ee_problem}
	\begin{align}
	\textbf{(EE)} \hspace{1em} & \max\sum_{k}\eta_k \\
	\text{s.t.} \hspace{1em} & R_k \geq R_k^{\min} \text{ for all } k \\
	& \sum_{k} \tau_k \leq 1 \label{ee_problem_st2}, \hspace{1em} 0 \leq \tau_k \leq 1 \text{ for all } k \\
	& N_{\text{T}}^{\min} \leq N_k^{\text{T}} \leq N_{\text{T}}^{\max} \text{ for all } k. \label{ee_problem_st4}
	\end{align}
\end{subequations}
The Lagrangian of (\ref{ee_problem}) can be written as
\begin{align}\label{lagrangian}
\mathcal{L}=\sum_{k}\eta_k+\sum_{k}\lambda_k\left(R_k-R_k^{\min}\right)+\mu \left(1-\sum_{k}\tau_k\right),
\end{align}
where $ \lambda_k $ is the Lagrangian variable associated with the minimum rate constraint of node $ k $ and $ \mu $ is the Lagrangian variable related to the access probability constraint. Another important objective that is widely used in resource allocation problems is the sum of the logarithm of energy efficiencies. To solve this problem, we replace the expression $ \sum_{k}\eta_k $ in (\ref{ee_problem}) and (\ref{lagrangian}) with $ \sum_{k}\log\left(\eta_k\right) $ and refer to the problem as \textbf{(LogEE)}. The \textbf{(LogEE)} problem trades off efficiency with fairness. The channel access probability and the PSDU frame length for all the nodes are represented by the vectors 
$\boldsymbol{\tau}=\left[\tau_1,\tau_2,...,\tau_{N_S}\right]$ and $\boldsymbol{N_{\text{T}}}=\left[N_1^{\text{T}},N_2^{\text{T}},...,N_{N_S}^{\text{T}}\right]$, respectively, 
where the symbols in bold define vectors. The optimal solution, $ \left(\boldsymbol{\tau^\star},\boldsymbol{N_{\text{T}}^\star}\right) $, as well as the corresponding Lagrangian variables $ \boldsymbol{\lambda^\star}=\left[\lambda_1^\star,\lambda_2^\star,...,\lambda_{N_S}^\star\right] $ and $ \mu^\star $ can be obtained by applying the Karush-Kuhn-Tucker (KKT) optimality conditions \cite{NonlinearBazaraa}.

\begin{algorithm}[t!] \label{algorithm}
	\caption{\small{EECAP: Energy Efficiency Optimization of Channel Access Probabilities}}\label{Algorithm}
	\small{{\begin{algorithmic}[1]
	\State Given $\boldsymbol{d},\boldsymbol{h},\boldsymbol{R^{\min}}$, and $N_S$
	\State Initialize $ \boldsymbol{\tau} $ and $ \boldsymbol{N_{\text{T}}} $
	\Repeat
	\For {node $ k=\{1,2,...,N_S\} $}
	\State solve (\ref{taukmin_THR}) to obtain $ \tau_{k,\min}^{\text{THR}} $
	\State solve (\ref{NkT_THR}) to obtain $N_{k,\text{T}}^{\text{THR}}$ 
	\EndFor
	\Until{stopping criteria is satisfied}
	\If{ $R_k\left(\boldsymbol{\tau}_{k,\min}^{\text{THR}},\textbf{N}_{k,\text{T}}^{\text{THR}} \right) \geq R_k^{\min} $ and $ \sum_{k}  \tau_{k,\min}^{\text{THR}} \leq 1 $ } \label{Algorithm:CheckCond}
	\Repeat \label{Algorithm:DualDecompStart}
	\For {node $ k=\{1,2,...,N_S\} $}
	\State solve (\ref{dualee_problem}) to obtain $ (\tau_k)^{\text{EE}} $ and $ (N_k^{\text{T}})^{\text{EE}} $
	\State $ \lambda_k=\max \left({R_k^{\min}-R_k\left(\boldsymbol{\tau^{\text{EE}}},\boldsymbol{N_{\text{T}}^{\text{EE}}}\right)},0\right)$ \label{Eqn:LambdaUpdate}
	\EndFor
	\State $ \mu^{\text{EE}} = \max \left( \sum_{k} (\tau_k)^{\text{EE}}-1,0 \right) \label{Eqn:MuUpdate}$ 
	\Until{stopping criteria is satisfied} 
	\State \textbf{return} $\left(\boldsymbol{\tau^\star},\boldsymbol{N_{\text{T}}^\star}\right)=\left(\boldsymbol{\tau^{\text{EE}}},\boldsymbol{N_{\text{T}}^{\text{EE}}}\right)$ \label{Algorithm:DualDecompEnd}
	\Else
	\Repeat \label{Algorithm:LogTHRStart}
	\For {node $ k=\{1,2,...,N_S\} $}
	\State solve (\ref{thr_problem}) to obtain $ (\tau_k)^{\text{LogTHR}} $ and $ (N_k^{\text{T}})^{\text{LogTHR}} $
	\EndFor
	\State $ \mu^{\text{LogTHR}} =\max ( \sum_{k} (\tau_k)^{\text{LogTHR}}-1,0 ) $
	\Until{stopping criteria is satisfied} 
	\State \textbf{return} $\left(\boldsymbol{\tau^\star},\boldsymbol{N_{\text{T}}^\star}\right)=\left(\boldsymbol{\tau^{\text{LogTHR}}},\boldsymbol{N_{\text{T}}^{\text{LogTHR}}}\right) $
    \label{Algorithm:LogTHREnd}
	\EndIf
\end{algorithmic}}}
\end{algorithm}
		
The problem (\ref{ee_problem}) is a single-ratio fractional program and it can be translated into a dual fractional program, which can be solved via the Gauss-Seidel iterative method \cite{NonlinearBazaraa}. The dual fractional program is given by \cite{HandbookCO}
\begin{align}\label{dualee_problem}
\textbf{(Dual-EE)} \qquad \min_{\boldsymbol{\lambda},\mu \geq 0}\left[\max_{\boldsymbol{\tau},\boldsymbol{N_{\text{T}}}}\mathcal{L}\right].
\end{align}
We denote this dual fractional program as \textbf{(Dual-EE)} if the maximization of the sum of energy efficiency is considered. If the sum of the logarithmic energy efficiency is evaluated, the corresponding Lagrangian will be used and the dual fractional program will be named as \textbf{(Dual-LogEE)}.

If there is no feasible solution for problem (\ref{ee_problem}), we solve the following problem \textbf{(LogTHR)}, which is defined as
\begin{align}\label{thr_problem}
\textbf{(LogTHR)} \qquad \max_{\boldsymbol{\tau},\boldsymbol{N_{\text{T}}}} \left(\sum_{k}\log R_k \right),
\end{align}
subject to (\ref{ee_problem_st2})-(\ref{ee_problem_st4}). The optimal solution 
$(\boldsymbol{\tau^{\text{LogTHR}}},\boldsymbol{N}_{\text{T}}^{\text{LogTHR}} ) $ is obtained by writing the Lagrangian of \textbf{(LogTHR)} and numerically solving for the solution. 

The proposed algorithm, EECAP, is presented under the heading Algorithm~\ref{Algorithm}. It consists of two stages. In the first stage, we check if the problem has a feasible solution. We solve (\ref{taukmin_THR}) and (\ref{NkT_THR}) to determine  $\tau_{k,\min}^{\text{THR}}$ and $N_{k,\text{T}}^{\text{THR}}$. In the next stage, if these values satisfy the check conditions (Step~\ref{Algorithm:CheckCond} of Algorithm~\ref{Algorithm}), then we solve (\ref{dualee_problem}) using dual decomposition method (Steps~\ref{Algorithm:DualDecompStart}-\ref{Algorithm:DualDecompEnd}). If the problem does not have a feasible solution that satisfies the rate and access constraints, then we solve the \textbf{(LogTHR)} problem that provides a fair solution (Steps~\ref{Algorithm:LogTHRStart}-\ref{Algorithm:LogTHREnd}). This typically occurs when the rate constraints are high or link distances are large.
We use the superscripts of $\mathrm{EE}$ and $\mathrm{LogTHR}$ above the variables to indicate the solutions of \textbf{(EE)} and \textbf{(LogTHR)} problems in Algorithm~\ref{Algorithm}, respectively. The dual decomposition method \cite{NonlinearBazaraa} is used to solve these two problems. Taking \textbf{(EE)} as an example, we first solve $ (\tau_k)^{\text{EE}} $ for node $ k $ via any numerical method such as the binary search while keeping the other parameters fixed. Then, we solve for $ (N_k^{\text{T}})^{\text{EE}} $ using the $ (\tau_k)^{\text{EE}} $ calculated in the previous step. Finally, the Lagrangian variable $ \lambda_k $ is updated in Step~\ref{Eqn:LambdaUpdate}. This process is repeated for $ N_S $ nodes. Next, the Lagrangian variable $ \mu^{\text{EE}} $ is updated in Step~\ref{Eqn:MuUpdate}. We repeat this procedure until the maximum number of iterations is reached or $ \left(\boldsymbol{\tau^{\text{EE}}},\boldsymbol{N_{\text{T}}^{\text{EE}}}\right) $ converges.

\begin{figure}[t!]
	\centering	
	  \subfigure{\includegraphics[width=0.5\columnwidth]{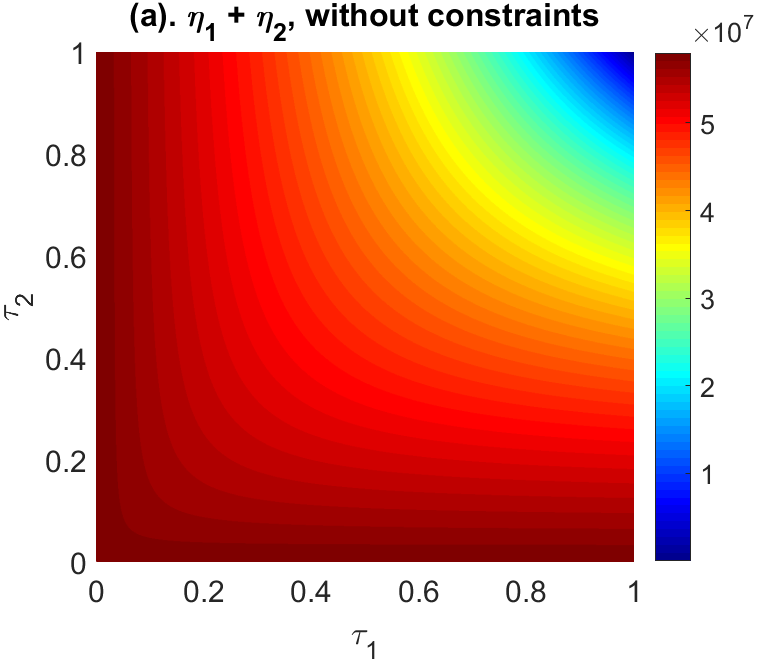}}\hspace{-0.15in}
	  \subfigure{\includegraphics[width=0.5\columnwidth]{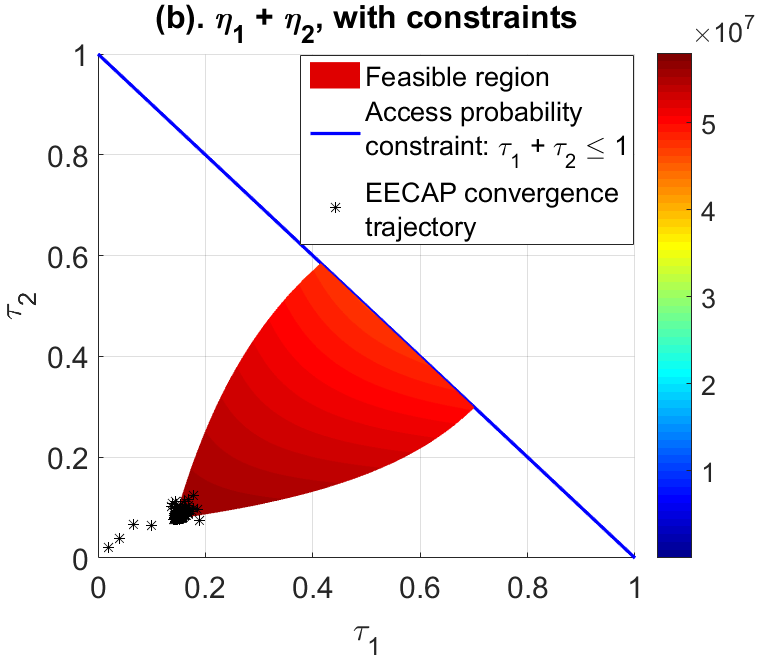}}
	  \subfigure{\includegraphics[width=0.5\columnwidth]{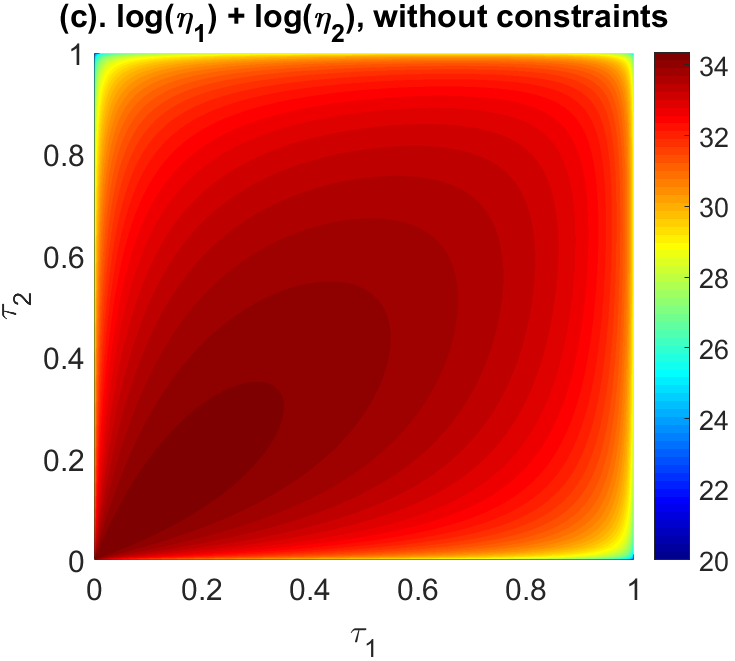}}\hspace{-0.15in}
	  \subfigure{\includegraphics[width=0.5\columnwidth]{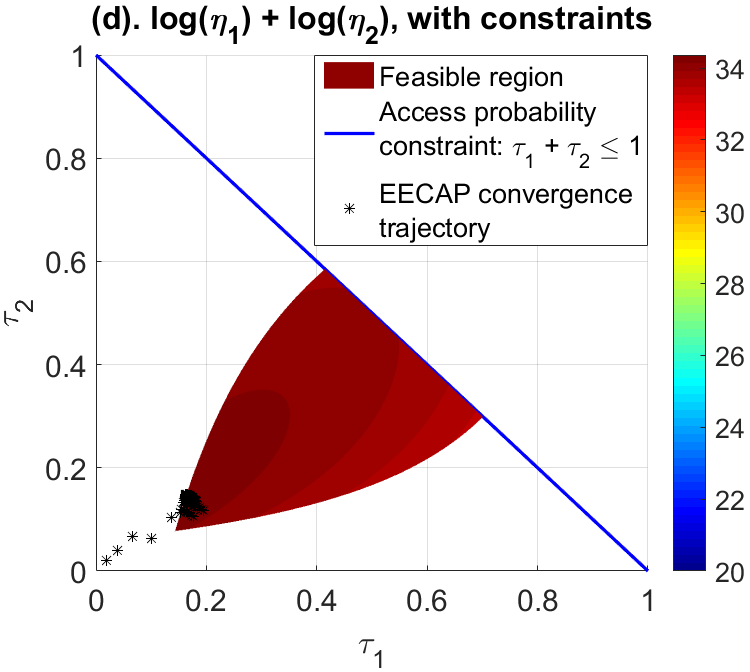}}
	\caption{Feasibility regions of the \textbf{(EE)} and \textbf{(LogEE)} problems with and without constraints are depicted. Link distances are 1 meter and the minimum rate constraint  taken as 1 Mbits/sec for node 1 and 0.5 Mbits/sec for node 2.}  
	\label{Fig_meshgrid}
\end{figure}

\begin{figure}[t!]
		\centering
		\includegraphics[width=0.7\columnwidth]{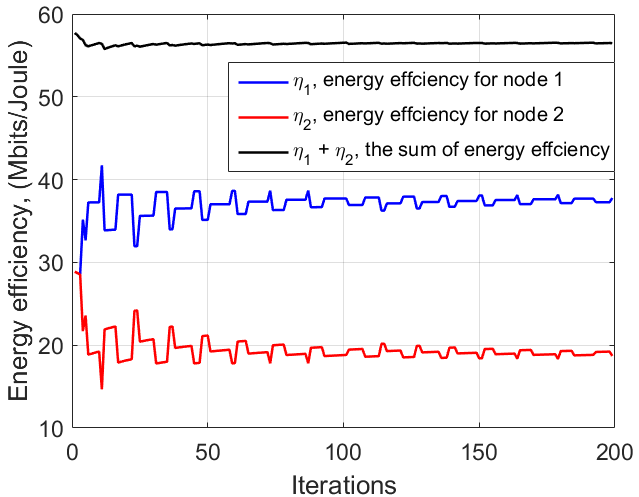}
		\caption{Energy efficiency versus iterations are shown to demonstrate the convergence of the solution of EECAP in Fig.~\ref{Fig_meshgrid}(b).}  
\label{Fig_iterations}
\end{figure}
	
%


\section{Simulation Results}\label{Section:Simulations}
In this section, we evaluate the performance of the proposed algorithm and provide insights on how the network parameters affect the solution. Simulations are conducted in MATLAB. The value of $ N_{\text{cpb}} $ is determined according to the link distances reported in Fig.~3(a) of \cite{CLOEE16}. We employ the UWB channel model for WBANs in \cite{BAN09}. The values of the parameters in the energy consumption model can be found in \cite{CLOEE16}.
  
In Figs.~\ref{Fig_meshgrid}(a)-(d), we first investigate the feasible regions of Problems \textbf{(EE)} and \textbf{(LogEE)} with and without rate constraints. For visual clarity, we present the results for only two nodes. Fig.~\ref{Fig_meshgrid}(a) depicts the sum of energy efficiencies under no constraints. It can be observed that the maximum energy efficiency is obtained when either node's access probability is zero. Similarly, the sum of logarithmic energy efficiencies is plotted in Fig.~\ref{Fig_meshgrid}(c), without any constraint. The maximum is at $ (0^+,0^+) $. The reason for this is that since the collisions waste energy and idle channel does not cost any energy, the system will try to avoid collisions at the cost of reduced access probability. Figs.~\ref{Fig_meshgrid}(b) and (d) illustrate the feasible regions with the rate and access constraints. We also plot the trajectory of the EECAP solutions with asterisk and show its evolution over 200 iterations. The link distance is set to be 1 meter for both nodes and $N_{\text{T}} = 2646$ bits. The rate constraint is set to be $ (R_1^{\min},R_2^{\min})=(1,0.5) $ Mbit/s. The optimal solution $ \boldsymbol{(\tau^\star)}^{\text{EE}} $ is $ (0.1430,0.0770) $ in Fig.~\ref{Fig_meshgrid}(b) and $ \boldsymbol{(\tau^\star)}^{\text{LogEE}}= (0.1610,0.1410) $ in Fig.~\ref{Fig_meshgrid}(d). We can see that for the Problem \textbf{(EE)}, $ \tau_1^\star $ is approximately twice as $ \tau_2^\star $, indicating that the solution of \textbf{(LogEE)} provides fairness between the two nodes, with closer $\tau_1^\star$ and $\tau_2^\star$ vales. For both cases, our algorithm is able to approach the optimal solution in a limited number of iterations. We show the rapid convergence of our proposed algorithm in Fig.~\ref{Fig_iterations} that depicts the energy efficiency versus iterations. Note that these are also the results shown in Fig.~\ref{Fig_meshgrid}(b). We set the initial value of the access probability as $ (0.01, 0.01) $ at which the sum of energy efficiencies is higher at first, but does not satisfy the rate constraints. After about 5 iterations, it steps into the feasible region and starts to look for the optimal solution that maximizes the sum of energy efficiencies. We can see that the fluctuations get smaller as more iterations are performed.


\begin{figure}[t!]
	\centering
	\begin{tabular}{cc}
		\subfigure[]{\includegraphics[width=0.46\columnwidth]{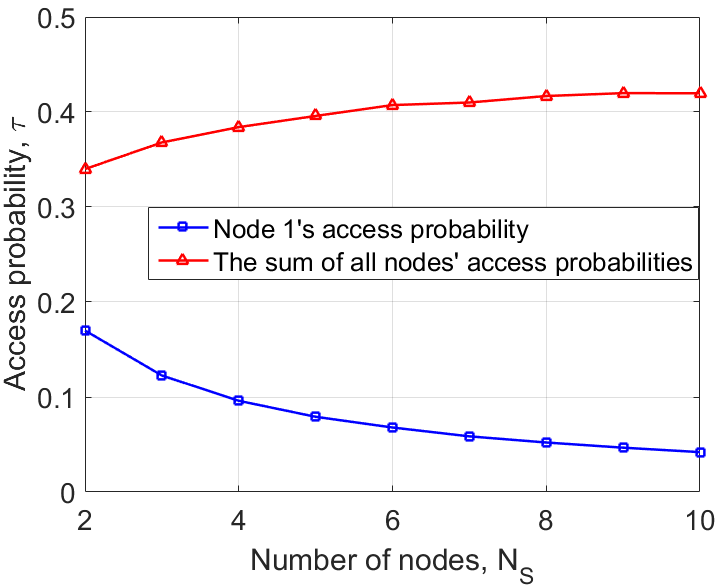}} &
		\subfigure[]{\includegraphics[width=0.46\columnwidth]{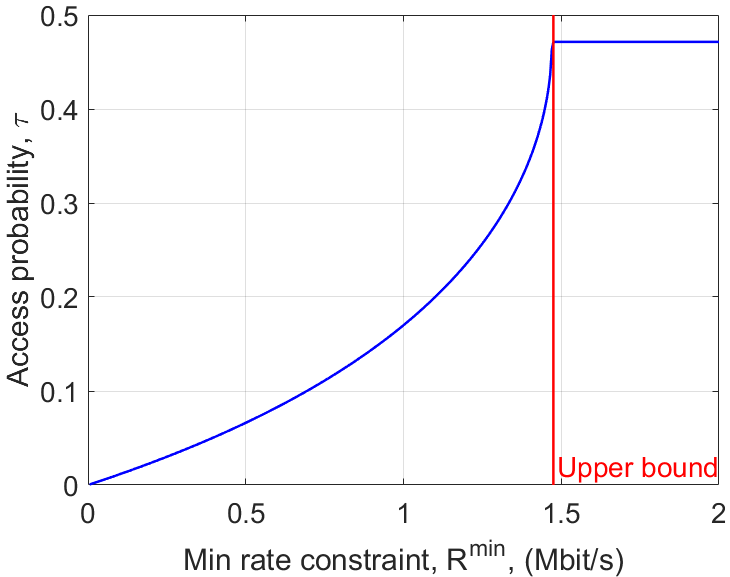}}			
	\end{tabular}
	\caption{\footnotesize{The optimal access probabilities versus the number of nodes (a) and different values of minimum rate constraints (b) are depicted. Link distance is 1 meter. In (a), the minimum rate constraint is taken as 1~Mbits/s. In (b), the problems \textbf{(Dual-EE)} and \textbf{(LogTHR)} are solved.}}  
	\label{Fig_combine}
\end{figure}

%
%
%

\begin{figure*}[t!]
	\centering
	\begin{tabular}{ccc}			
		\subfigure[]{\includegraphics[width=0.31\textwidth]{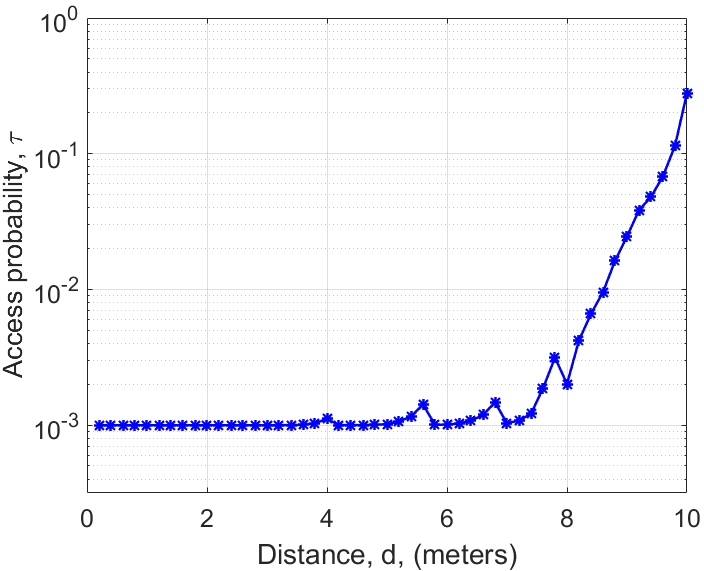}} & 
		\subfigure[]{\includegraphics[width=0.31\textwidth]{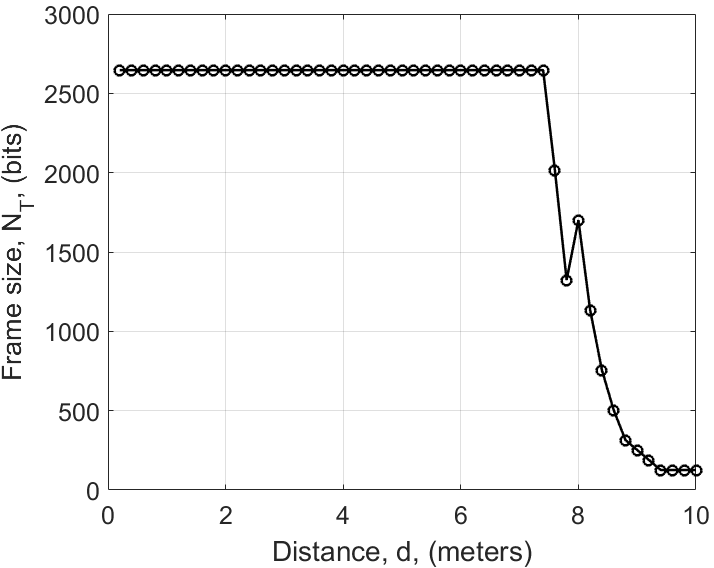}} &
		\subfigure[]{\includegraphics[width=0.31\textwidth]{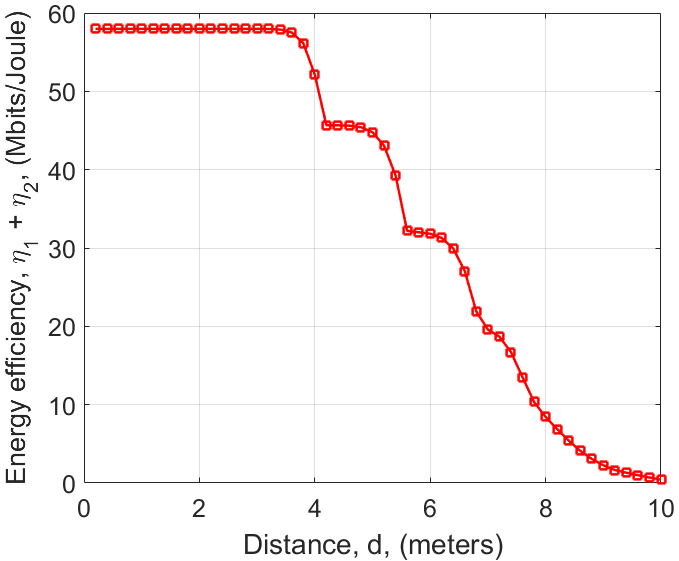}}			
	\end{tabular}
	\caption{Link adaptation results of the proposed algorithm. Rate constraint is taken as 1~Mbit/sec for two nodes.}  
	\label{Fig_distance}
\end{figure*}

Fig.~\ref{Fig_combine}(a) depicts the effect of the number of nodes on the optimal channel access probability that solve \textbf{(EE)}. The rate constraint is fixed to 1~Mbits/sec. The link distances are taken as 1~meter and we increase the number of nodes from two to ten. For an individual node, its optimal access probability decreases as the number of nodes grows. However, when we consider the overall system utilization, defined as the sum of individual access probabilities, it increases with the number of nodes. Fig.~\ref{Fig_combine}(b) shows the optimal access probability versus the rate constraints. There are two nodes with link distances of 1~meter. We note the upper bound of the minimum rate constraint is 1.475~Mbits/s. There is no feasible solution beyond this limit and \textbf{(LogTHR)} is solved to find the optimal access probability that maximizes the sum of logarithmic throughputs. Before reaching this limit,  the optimal access probability follows an exponential increase with the rate constraint.

The distance between the node and the hub has a strong impact on the optimal channel access probability and frame length, since the probability of successful delivery of frames drops as the distance increases. We depict the distance versus the optimal access probability, optimal PSDU frame size, and sum of energy efficiencies in Figs.~\ref{Fig_distance}(a)-(c), respectively. There are two nodes and the rate constraint is fixed to 18 Kbps. We only show the optimal access probability of node 1 as the probabilities are identical for this setting. For link distances above 7.5 meters, the error probability starts to increase significantly. This results in a sudden increase of the optimal access probability and a steep decrease of the optimized frame size. The same observation for the frame size was also reported in our prior work \cite{CLOEE16}. The maximum energy efficiency also drops with distance. The transitions in Figs.~\ref{Fig_distance}(a)-(c) reflect the points where $ N_{\text{cpb}} $ changes. For example, at the link distances from $ d=7.8 $ m to $ d=8 $ m, the decrease in the optimal access probability (see Fig.~\ref{Fig_distance}(a)) and the increase in frame length (see Fig.~\ref{Fig_distance}(b)) are due to the selection of a higher $ N_{\text{cpb}} $ value to increase robustness.

\section{Conclusion}\label{Section:Conclusion}
In this paper, we address the energy efficiency maximization problem under the rate and access constraints for one-hop star WBANs. We derive expressions of the PHY and MAC layer successful transmission probabilities. We propose an algorithm that determines optimal channel access probabilities and frame sizes for each node. We also study two different energy efficiency models, \textbf{(EE)} and \textbf{(LogEE)} that emphasize the efficiency and fairness trade-off. The performance of the EECAP algorithm is evaluated with simulations. Our results demonstrate that the optimal access probability increases with the link distance, increases exponentially with the minimum rate constraint, and decreases logarithmically with the number of nodes. In future work, the proposed algorithm will be extended to two-hop WBANs and Internet of Things (IoT) networks. Also, we plan use Game Theory to manage other network problems such as nodes not honoring their service agreements. 


\appendix\label{Appendix}
The derivative of $ R_k $ respect to $ \tau_k $ is,
\begin{align}
\frac{\partial R_k}{\partial \tau_k}=\frac{N_k^{\text{T}}(1-p_k)P_k^{\text{SHR}}P_k^{\text{PHR}}(P_k^{\text{CW}})^{\frac{N_k^{\text{T}}}{n}} \cdot YT}{\left(XT\cdot\tau_k+YT\right)^2},
\end{align}
where $YT$ is as defined in Section~\ref{Section:Formulation}. Since $ T_k^{\text{S}} > T_k^{\text{C}} $  \cite{Cagalj05},
\begin{align}\begin{aligned}
YT=&y_k^{\text{S}}T_k^{\text{S}}+y_k^{\text{C}}T_k^{\text{C}}+y_k^{\text{I}}T_k^{\text{I}}\\ 
=&x_k^{\text{C}}T_k^{\text{S}}+(p_k-x_k^{\text{C}})T_k^{\text{C}}+(1-p_k)T_k^{\text{I}} \geq 0.
\end{aligned}\end{align}
Therefore $ \partial R_k/\partial \tau_k \geq 0 $.

\section*{Acknowledgment}
Yang Liu's participation in this publication was made possible by NPRP grant \#6-415-3-111 from the Qatar National Research Fund (a member of Qatar Foundation). The statements made herein are solely the responsibility of the authors.

\bibliographystyle{IEEEtran}
\bibliography{IEEEabrv,AdaptiveFFR}

\end{document}